\documentclass{article}
\usepackage{bm}
\usepackage[T2A]{fontenc}
\usepackage[utf8]{inputenc}
\usepackage[english]{babel}
\usepackage{amsmath,amssymb,amsfonts}
\usepackage[a5paper,nohead,headsep=0.6cm,left=1.5cm,
right=1cm,top=2cm,bottom=2cm,twoside]{geometry}
\usepackage{graphicx}

\newtheorem{theorem}{Theorem}
\newtheorem{corollary}{Corollary}

\newtheorem{lemma}{Lemma}

\newtheorem{definition}{Definition}

\begin{document}

\begin{center}
	\title{}{\bf A UNIQUENESS RESULT FOR THE INVERSE PROBLEM OF IDENTIFYING
		BOUNDARIES FROM WEIGHTED RADON TRANSFORM}
	
	\author{}{D.S. Anikonov}, {S.G. Kazantsev}, {D.S. Konovalova}
	
Sobolev Institute of Mathematics,  Novosibirsk, Russia

e-mail:anik@math.nsc.ru, kazan@math.nsc.ru, dsk@math.nsc.ru

\end{center}

 \begin{quote}
	\noindent{\bf Abstract. } We study the problem of the integral geometry, in which
	the functions
	are integrated over hyperplanes in the $n$-dimensional Euclidean
	space, $n=2m+1$. The integrand is
	the product of a function of $n$ variables called the density and
	weight function depending on $2n$ variables. Such an integration
	is called here the weighted  Radon transform,
	which coincides with the classical one if the weight function is equal to
	one. It is proved the uniqueness for the problem of
	determination of the surface on which the  integrand is
	discontinuous.
	
\end{quote}

\section{Introduction}
In a broad sense, the problem of the integral geometry consists in
    obtaining information about the integrand by values
    some set of integrals. For some versions of the Radon
    transform,      the uniqueness of the solution is proved for the problem of determining
    integrand under fairly general assumptions. However
    obtaining constructive formulas required much more
    strict restrictions such as smoothness integrand.

    The specifics of this work
    consists, in particular, in the use of piecewise continuous
    integrands. Another feature of the study is
    in the fact that under the integral there is an unknown function depending
    on $2n$ variables, and the set of known integrals depends on
    $2n-1$  variables.  This ratio of dimensions does not allow
    hope to find the entire integrand. That's why
    another problem is posed of determining only the discontinuity surfaces
    integrand. Two theorems of the uniqueness are proved and
    the corresponding algorithm is proposed.

 We emphasize  that knowledge of the density discontinuity
surfaces is a useful information. For example, in probing problems
such information is very important. In particular, for X-ray
tomography, this information is basic. Note that the topic of
searching for unknown boundaries is quite extensive and refers to
different areas of mathematics. Probably such a first setting is
Stefan's famous problem.

The present work is a continuation previous research of the
authors in the field of the integral geometry \cite{ABK:2022},
\cite{AK:2015}, \cite {APK:1993}. Without claiming to be a
complete review of the topic, we point to the works of the
classics of this field, such as J. Radon \cite{Hel:1999}, R.
Courant \cite{Courant:1964}, F. John \cite{John:1958}, and M.
Gelfand \cite{GGV:1966}. In addition, a significant contribution
to the solution of such problems was given  by the works of the
mathematical school of M. M. Lavrent'ev and V.G. Romanov devoted
to the study of inverse problems of mathematical physics
\cite{LavSav:2006}.

Currently, research in this direction continues, albeit with less
intensity. In particular, we can specify the publications \cite
{DMS:2018}, \cite{KIA:2021}, \cite{Koganov:2011}, \cite{LLR:2019},
\cite {M:2006}, \cite{Natterer:1986}, \cite {SPA:2019},  \cite
{TK:2020},  \cite {VKF:1985}.
  Of these, we note the works \cite
{DMS:2018}, where the numerical experiments were performed for the
problems of the integral geometry with discontinuous integrands.

\section{Basic notation and definitions}
Consider an $n$-dimensional Euclidean space $E_n,\ n \geq 3$, in
which the orthonormal basis ${\bf e}_1,...,{\bf e}_n$ is given. The default
for the coordinate representation of vectors from $E_n$ this basis
will be used. The use of other coordinates will be specially
stipulated.

Let us introduce the following notations: $B({\bf x},\delta)=\{
{\bf y}: {\bf y}\in E_n, |{\bf y}-{\bf x}|<\delta, \ {\bf x}\in
E_n \}$; $\partial T$ -- the boundary of the set $T$; ${\Omega}$
is the unit sphere in $E_n$; const -- some positive number;
$\Delta_{\bf x}$-- the Laplace operator with respect to the
variable ${\bf x}=(x_1,...,x_n)$; $C^k(T)$-- the space of
functions defined on the set $T$ and having all continuous
derivatives up to the $k$-th order inclusive; $\mu_N(T)$ -- the
$N$-dimensional Lebesgue measure of the set $T$; $\mu_{\Omega}(Q)$
is the Lebesgue measure of the subset $Q\subset {\Omega} $ by the
measure defined on the sphere $\Omega$.

Note that for the Radon transform, the cases of even and odd
dimensions differ quite significantly. In this work we
consider the variant of the odd $n=2m+1, \ m=1,2,...$\ .

Let $G$ be a bounded domain in $E_n$ containing pairwise disjoint
domains $G_i, \ i=1,...,l$. Denoting the union of these sets by
$G_0$, we require that $\overline{G}_0 =\overline{G}$. Each
boundary $\partial G_i$ is assumed to be an $(n-1)$-dimensional
continuous closed surface. It's clear that the surface $\partial
G_0$ coincides with the union of the surfaces $\partial G_i, \
i=1,...,l$.

\begin{figure}[!b]\centering
\includegraphics[width=0.5\textwidth,height=12pc]{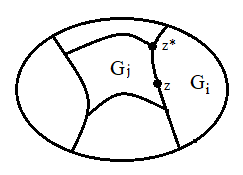}
\caption{Two-dimensional  section of the set $G_0$. A point $z$
is a contact  and $z^*$ is not a contact point. \label{fig} }
\end{figure}

A point $ {\bf z}\in \partial G_0$ is called a contact point if it
belongs to a common segment of the boundary of exactly two sets
$G_i$ and $G_j$, $1\leq i,j \leq l$, (see Fig \ref{fig}). We will
assume that the set contact points is dense in $\partial G_0
\backslash
\partial G$ .

Let us define the following class of functions $K$. The bounded
 function $v({\bf y}), \ {\bf y}\in E_n$ belongs to the class $K$,
if the following conditions hold
$$|v({\bf y})-v( \tilde{{\bf y}})|\leq const |{\bf y}- \tilde{{\bf y}}|^{\alpha}, \ {\bf y},
\tilde{{\bf y}}\in G_i, \ i=1,...,l,$$ $ \ 0 <\alpha \leq 1$ and $v({\bf y})=0$
for ${\bf y} \notin G$.

In other words, we consider Holder piecewise continuous supported
functions in $\overline G$. It is clear that for $v({\bf y})\in K$
and for each $i, \ 1\leq i\leq l$, there are the limit finite
boundary values $[v({\bf z})]_i$, ${\bf z} \in \partial G_i$, i.e.
$v({\bf y}) \to [v({\bf z})]_i $ for ${\bf y}\to {\bf z},\ {\bf
y}\in G_i$.

At the contact points ${\bf z}\in \partial G_0$ we define the
jumps of the function by the equalities $[v({\bf z})]_{
    i,j}=[v({\bf z})]_{j}-[v({\bf z})]_{i}$, $1\leq i<j \leq l$, ${\bf z}\in
\partial G_i\cap \partial G_j$.

We introduce the following
\begin{definition}
Let $v({\bf y})\in K$, $V({\bf x},{\bf y})\in C^{2m}(E_n \times
E_n)$. Let's define weighted  Radon transform by the formula
\begin{equation}\label{1}
[{\mathcal U}v]({\bf x},{\bm \omega})=\int_{({\bf y}-{\bf x}) \centerdot
{\bm \omega}=0}V({\bf x},{\bf y})v({\bf y})d_{\bf y}\sigma, \
{\bf x}\in G, \ {\bm \omega} \in  \Omega \ .
\end{equation}
    \end{definition}

Here integration is performed over the hyperplanes $Y({\bf x},{\bm
\omega})=\{{\bf y}:{\bf y} \in E_n, \ ({\bf y}-{\bf x}) \centerdot
{\bm \omega} =0 \} $. It is clear, that the restrictions on the
functions $V({\bf x},{\bf y})$ and $v({\bf y})$ are sufficient for
the existence of the integral on the right side of the formula
(\ref{1}). For the same hyperplanes, we will also use another,
more common notation $L({ \bm \omega}, p)=\{{\bf y}:{\bf y} \in
E_n, \ {\bf y} \centerdot {\bm \omega} =p \} $. It is not hard to
see that $L({\bm \omega}, {\bf x} \centerdot {\bm \omega})=Y({\bf
x},{ \bm \omega})$. Following some traditions, we will call the
function $V({\bf x},{\bf y})$ weight, and the function $v({\bf
y})$ is the  density.

We write the classical Radon transform in the form
$$[{\mathcal R}]({\bm \omega},p)=\int_{{\bf y} \centerdot { { \bm \omega}}=p}v({\bf y})d_{\bf y}\sigma,
\  { \bm \omega} \in {\Omega}, \ -\infty <p<\infty .
$$

The number $p$ is the deviation of the hyperplane $L({\bm \omega}, p)$
from the origin. It is easy to see that if $V({\bf x},{\bf y})\equiv 1$, then
$[{\mathcal U} v]({\bf x},{\bm \omega})=[{\mathcal R }v]({\bm \omega},{\bf x} \centerdot{\bm \omega})$.

We accept one more additional restriction for the set $G_0$. We
will say that $G_0$ is a generalized convex set if the following
assumptions hold.

There is a subset ${\Omega}^{\prime}$ of the sphere ${\Omega}$,
different from $\Omega$ by a set of a zero measure  with the next
conditions.

1) For any vector ${ { \bm \omega}} \in { \Omega}^{\prime}$ the numbers
$p^{+}({ { \bm \omega}})$ and $p^{-}({ { \bm \omega}})$ are defined such that for
$p\geq p^{+}({ { \bm \omega}})$ and for $p\leq p^{-}({ { \bm \omega}})$ the  property
$L({ { \bm \omega}},p)\cap G = \varnothing$ is valid.

2) For $p$, $p^{-}({ { \bm \omega}})<p<p^{+}({ { \bm \omega}})$
the hyperplane $L({ { \bm \omega}},p)$ intersects the set $G_0$,
and the boundary of the intersection $G_0^{\prime}=L({ { \bm
\omega}},p)\cap G_0$ has a zero measure, i.e.,
$\mu_{n-1}\partial(L({ { \bm \omega}},p)\cap G_0))=0$.

3) If $p\to p^{+}({ { \bm \omega}})$ or $p\to p^{-}({ { \bm \omega}})$ then
$\mu_{n-1}(L({ { \bm \omega}},p)\cap G_0)\to 0$.

The above definition makes it possible to include a significant number of cases in the number of cases under study. For clarification, we give two simple but typical examples.

{\bf Example 1.} Let $G$ be the ball $B({\bf 0},2\delta)$,
$G_1=\{{\bf y}: \in E_n, -\delta <y_i <\delta, i=1,...,n \}$ --
the cube in $E_n$, $G_2=B({\bf 0},2\delta)\backslash \overline
G_1$, $G_0=G_1\cup G_2$. In this case  ${\bm \Omega}^{\prime}$
does not contain only the vectors ${\bf e}_1,...,{\bf e_n}$,
$-{\bf e}_1,...,-{\bf e_n}$. Thus  all the generalized convexity
requirements are  satisfied.

{\bf Example 2.} Let $G=B_1\cup B_2$, where $B_1$ and $B_2$ are
two balls in $E_n$ at a positive distance from each other. Then
for $G_0=B_1\cup B_2$ the condition of item 2 of generalized
convexity is not satisfied. However, if we add another ball $B_3$
containing $B_1\cup B_2$, then for $G_0=B_1\cup B_2\cup
(B_3\setminus \overline{B_1\cup B_2} )$ all generalized convexity
conditions are satisfied, and ${\Omega}^{\prime}=\Omega$.

\section{Statement and study of the problem}

Using the previous notation and definitions, consider
the following problem about the unknown boundary.

{\bf Problem.} Given the domain $G$ and the function $[{\mathcal U}v]({\bf x},{\bm \omega})$,
    ${\bf x}\in G, {\bm \omega} \in {\Omega}, \ v\in K$ find the surface
$\partial G_0$.

Now  we present two auxiliary statements.

\begin{lemma} If $\lambda({\bf y})$ belongs to the class $K$, then for
any vector ${ { \bm \omega}} \in {\Omega}^\prime$ the Radon
transform $[{\mathcal R}\lambda] ({\bm \omega},p)$ is continuous
in $p$.
\end{lemma}
{Proof.}
We consider an arbitrary vector ${\bm \omega} \in {\Omega}^\prime$
and use the Cartesian coordinate system in which the orts of the
coordinate axes are ${\bf e}_1^{\prime},...,{\bf e}_n^{\prime}$,
where ${\bf e}_n^{\prime}={\bm \omega}$. In this coordinate
system, the Radon transform  takes on a simple form
$$[{\mathcal R}\lambda] ({\bf e}_n^{\prime},p)=\int \lambda(\xi_1,...,\xi_{n-1},p)d\,\xi_1 \,...d \,\xi_{n-1}.
$$
We use the numbers $p^{+}({\bf e}_n^{\prime})$ and $p^{-}({\bf
e}_n^{\prime})$, specified in the definition of the generalized
convexity. First consider the case $p^{-}({\bf
e}_n^{\prime})<p<p^{+}({\bf e}_n^{\prime})$. Such the value $p$
corresponds to the section $L({\bm e}_n^{\prime},p)\cap G_0$.
Let's take an arbitrary point ${\bm \xi}=(\xi_1,...,\xi_{n-1},p)$
from this section.
Since ${\bm \xi}$ also belongs to the set $G_0$, then there exists
a ball $B({\bm \xi}, \delta)\subset G_0 $ in which the function
$\lambda$ is continuous.
Let $\{ p_k \}$ be an arbitrary sequence converging to $p$. It is
clear that starting from some number $k$ all points
$(\xi_1,...,\xi_{n-1},p_k) \in B({\bm \xi}, \delta)$. Therefore
$\lambda (\xi_1,...,\xi_{n-1},p_k)$
$\to \lambda
(\xi_1,\dots,\xi_{n-1},p)$ for $k\to \infty$. By virtue of the
choice point $(\xi_1,\dots,\xi_{n-1},p)$ such convergence takes
place for almost all $\xi_1,\dots,\xi_{n-1}$ corresponding to the
section $L({\bm e}_n^{\prime},p)\cap G$.

Also we note that  the functions
$\lambda(\xi_1,...,\xi_{n-1},p_k)$ are bounded by the single
constant. Hence, by the Lebesgue's dominated convergence theorem
we derive
$$ \lim_{k\to \infty}\int
\lambda(\xi_1,...,\xi_{n-1},p_k)d\,\xi_1 \,...d \,\xi_{n-1}= \int
\lambda(\xi_1,...,\xi_{n-1},p)d\,\xi_1 \,...d \,\xi_{n-1} ,
$$
which means that $[{\mathcal R}\lambda]({ { \bm \omega}},p)$ is
continuous for $p^{-}({ { \bm \omega}})<p<p^{+}({ { \bm
\omega}})$. As for the continuity $[{\mathcal R}\lambda]({ { \bm
\omega}},p)$ for $p \leq p^{-}({ { \bm \omega}})$ and $p \geq
p^{+}({ { \bm \omega}})$, then it is a simple consequence of the
restrictions of the first and third points of the generalized
convexity conditions. The lemma is proven.
\begin{lemma}
 If
$\lambda({\bf y}) \in K$ then the identity is valid
\begin{equation}\label{2}
\int_{{ { \Omega}}}[{\mathcal R}\lambda] ({ { \bm \omega}}, {\bf
x}\centerdot { { \bm \omega}})d{ { \bm
\omega}}=\frac{2m\pi^{m}}{m!}\int_{G}\frac{\lambda({\bf y})}{|{\bf
y}-{\bf x}|}d{\bf y}.
\end{equation}
\end{lemma}
{Proof.} 
	 Following the  scheme used in \cite{John:1958} for
smooth functions, consider the expression
$$W({\bf x})=\int_{G}\int_{\Omega}\lambda({\bf y})|({\bf y}-{\bf x})\centerdot { { \bm \omega}}|)d{ { \bm \omega}} d{\bf y}
$$
and express it in terms of the Radon transform. Namely, by changing the order of integration and using the Fubini theorem, we obtain

$$W({\bf x})=\int_{\Omega}\int_{G}\lambda({\bf y})|({\bf y}-{\bf x}) \centerdot{ { \bm \omega}}|)d {\bf y} \,d{ { \bm \omega}}=
\int_{\Omega}\int_{-\infty}^{\infty }|p|\int_{({\bf y}-{\bf x}) \centerdot { { \bm \omega}}=p}
\lambda({\bf y}) d_{\bf y}\sigma \,dp\,d{\bm \omega}
$$
$$=
\int_{\Omega}\int_{-\infty}^{\infty
}|p|[{\mathcal R}\lambda]({ { \bm \omega}},p+{\bf x}\centerdot {\bm \omega})
\,dp\,d{\bm \omega}.
$$

Consider the inner integral over $p$ for any fixed ${\bm \omega} \in
{\Omega}^{\prime}$. Then according to lemma 1 $[{\mathcal R} \lambda] ({\bm \omega},
p+{\bf x}\centerdot { { \bm \omega}})$ is a continuous function in variable $p$.
This property, as well as the obvious identity $[{\mathcal R}\lambda]
({\bm \omega}, p)=[{\mathcal R}\lambda] (-{\bm \omega}, -p)$ allow us to perform the
following transformations
$$\Delta_{\bf x} \int_{-\infty}^{\infty }|p|[{\mathcal R}\lambda({ { \bm \omega}},p+{\bf x}\centerdot
{ { \bm \omega}})]
\,dp
$$
$$
= \Delta_{\bf x}\left [\int_{{\bf x}\centerdot { { \bm \omega}}}^{\infty
}(p-{\bf x}\centerdot { { \bm \omega}})[{\mathcal R}\lambda]({ { \bm \omega}},p) \,dp
- \int_{-\infty}^{{\bf x}\centerdot {\bm \omega} }(p-{\bf x}\centerdot { { \bm \omega}})[{\mathcal R}\lambda]({\bm \omega},p)
\,dp \right ]
=2[{\mathcal R}\lambda] ({ \bm \omega},{\bf x}\centerdot { { \bm \omega}}).
$$

Hence,
\begin{equation}\label{3}
\Delta_{\bf x}W({\bf x})=2\int_{\Omega }
[{\mathcal R}\lambda] ({\bm \omega},{\bf x}\centerdot {\bm \omega})d {\bm \omega} .
\end{equation}

Here and below, we use the fact that integrations over ${\Omega}$
and ${\Omega}^{\prime}$ are equivalent.

The value of $\Delta_{\bf x}W({\bf x})$ can also be calculated in another way.
 To do this, we use the well--known identities \cite{John:1958}

\begin{align}\label{4}
\int_{{ { \Omega}}}|\bm \eta \centerdot { { \bm \omega}} |d { {
\bm \omega}}&=\frac{2\pi^m}{m!}|\bm \eta|, \ \bm \eta \in E_n,
\\ \label{44}
\Delta_{\bf x}|{\bf y}-{\bf x}|&=2m|{\bf y}-{\bf x}|^{-1}.
\end{align}

Note that equation (\ref{4}) can easily be deduced from the
Funk--Hecke theorem.

Thus we can write
$$W({\bf x})=\frac{2\pi^m}{m!}\int_{G}\lambda({\bf y})|{\bf y}-{\bf x}|d{\bf y},
$$

\begin{equation}\label{5}
\Delta_{\bf x}W({\bf x})=\frac{4m\pi^m}{m!}\int_{G}\frac{\lambda({\bf y})}{|{\bf y}-{\bf x}|}d{\bf y}.
\end{equation}

Comparing (\ref{3}) and (\ref{5}) we get (\ref{2}). The lemma is
proved.

We remark that the statement of this lemma is a generalization of
the result from Natterer's book \cite{Natterer:1986}.

\begin{corollary}
    Let $v({\bf y})\in K$ then  the equality takes place
\begin{equation}
\label{6}
\int_{\Omega}[{\mathcal U}
v]({\bf x},{ { \bm \omega}})\,d{ { \bm \omega}}=\beta_1\int_{G}\frac{V({\bf x},{\bf y})v({\bf y})}{|{\bf y}-{\bf x}|}d{\bf y}, \ \beta_1=\frac{2m\pi^m}{m!}.
\end{equation}
\end{corollary}

{Proof.}   Note that in the equality (\ref{2}) the integration
is produced over ${\bf y}$ and over ${ { \bm \omega}}$, while the
variable $x$ unchanged. This allows us to consider ${\bf x}$ as a
parameter here, which expands the possibilities of applying the
formula (\ref{2}). Namely, for a fixed value of ${\bf x}$, we can
consider the product $V({\bf x},{\bf y})v({\bf y})$ as some
function $\lambda({\bf y})$ and assume that $\lambda({\bf y})\in
K$ and $[{\mathcal U}v]({\bf x},{ { \bm \omega}})= [{ \mathcal
R}\lambda] ({\bm \omega}, {\bf x} \centerdot{ { \bm \omega}})$.
Then the equality (\ref{2}) implies the equality (\ref{6}), which
proves the corollary 1.

\begin{theorem}
For any function $v({\bf y})\in K$ the following equality is valid
\begin{equation}\label{7}
(\Delta_{\bf x})^m\int_{\Omega}[{\mathcal U}
v]({\bf x},{ \bm \omega})d{ { \bm \omega}}=2(2\pi)^{2m}(-1)^mV({\bf x},{\bf x})v({\bf x}) +\Phi({\bf x}),
\ {\bf x}\in G_0,
\end{equation}
where $\Phi({\bf x})$ is a continuous and bounded function for ${\bf x}\in G$.
\end{theorem}
{Proof.}
In addition to the identities (\ref{4}), (\ref{44}) we will use
also the following equalities given, for example, in
\cite{John:1958}

\begin{align}\label{8}
(\Delta_x)^m|{\bf y}-{\bf x}|&=\beta_2|{\bf y}-{\bf x}|^{2-n},
\\ \label{88}
\Delta_x\int_{G}\frac{v({\bf y})}{|{\bf y}-{\bf x}|^{n-2}}d{\bf
y}&=\beta_3v({\bf x}), \ {\bf x} \in G_0,
\end{align}
where
\begin{align}\label{9}
\beta_2&=(-1)^m\frac{2^n\Gamma(3/2)\Gamma(m+1)\Gamma(n/2)}{\pi(2-n)},
\\ \label{99}
\beta_3&=\frac{2 \pi^{n/2} (2-n)}{\Gamma (n/2)}.
\end{align}

Denote the integral on the right side of the formula (\ref{6}) by $H({\bf x})$ and represent it as a sum:
$H({\bf x})=H_1({\bf x})+H_2({\bf x})$, where
\begin{align*}H_1({\bf x})&=\int_{G}V({\bf x},{\bf x})\frac{v({\bf y})}{|{\bf y}-{\bf x}| }d{\bf y},
\\
H_2({\bf x})&=\int_{G}(V({\bf x},{\bf y})-V({\bf x},{\bf x}))\frac{v({\bf y})}{|{\bf y}-{\bf x}| }d{\bf y}.
\end{align*}

It is easy to see that $ (\Delta_{\bf x})^m H_2({\bf x}) $ is a
linear combination of integrals of the potential type with weak
singularities
$$\Psi({\bf
x})=\int_{G_i}{\psi({\bf x},{\bf y})}\,d{\bf y},$$ where
$\psi({\bf x},{\bf y})$ is a continuous function for $(x,y) \in G
\times G_i$, $x \neq y$, $i = 1,...,l$ and $|\psi({\bf x},{\bf
y})| \leq const |y-x|^{1-n}$.

The properties of such integrals are well known. In particular, we
can refer to the theorem 1.1.1 from \cite{Mikhlin:1977} and state
that the function $\Psi({\bf x})$ is bounded and continuous in
$G$. Therefore  the function $(\Delta_{\bf x})^m H_2({\bf x}), \
{\bf x} \in G $ is also  bounded and continuous in $G$.

Now consider the expression $(\Delta_{\bf x})^m H_1({\bf x})$. We
use the equality $ (\Delta_{\bf x})^m H_1({\bf x})=h_1({\bf
x})+h_2({\bf x}) $, where
\begin{equation*}
h_1({\bf x})=\frac{V({\bf x},{\bf x})}{2m}(\Delta_{\bf x})^m\int_{G}\Delta_{\bf x}v({\bf y})|{\bf y}-{\bf x}|
\, d{\bf y}.
\end{equation*}

The term $h_2({\bf x})$ contains all derivatives, except those
distinguished for $h_1({\bf x})$. From this we can conclude that
the term $h_2({\bf x}), \ {\bf x}\in G$ is also a linear
combination of integrals of the $\Psi({\bf x})$ type and is also a
continuous and bounded function. For the transformation $h_1({\bf
x})$ we use the identities (\ref{8}), (\ref{88}) and write for
${\bf x}\in G_0$

\begin{equation*}
h_1({\bf x})=\frac{V({\bf x},{\bf x})}{2m}\Delta_x\int_{G}\beta_2\frac{v({\bf y})}{|{\bf y}-{\bf x}|^{n-2}}\ d
{\bf y}=\frac{\beta_2\beta_3}{2m}V({\bf x},{\bf x})v({\bf x}).
\end{equation*}

Due to (\ref{9}), (\ref{99})  it is easy to check that
$\beta_1\beta_2\beta_3/2m=2(-1)^m(2\pi)^{2m}$. Then, to derive the
equality (\ref{7}), it suffices to take into account the formula
(\ref{6}) and the equality
\begin{equation*}
\beta_1(\Delta_{\bf x})^{m}H({\bf x})=\beta_1h_1({\bf x})+F({\bf x}),
\end{equation*}
where $F({\bf x})$ is a continuous and bounded function for ${\bf x}\in G$. Thus, the theorem is proved.

     Note that if $V({\bf x},{\bf x})\neq 0$, then the right-hand side of the formula (\ref{7}) is
 a discontinuous function at points of the surface $\partial G_0$ only.

    We now turn to the formulation and proofs of the theorems
uniqueness of the problem. We take two sets of the elements
$V_i({\bf x},{\bf y})$, $v_i({\bf x})$, $\partial G^{(i)}_0$ and
the weighted Radon transforms $[{\mathcal U} v_i]({\bf x},{\bm
\omega})$, $i=1,2$.

\begin{theorem}
If at the contact point ${\bf z}^{(i)} \in \partial G_0^{(i)}$
$V_i({\bf z}^{(i)},{\bf z}^{(i)})\neq 0$ and functions $v_1, \
v_2$ have non-zero jumps at the points ${\bf z}^{(1)}$ and ${\bf
z}^{(2)}$, respectively , then from the match $[{\mathcal U} v_1]
({\bf x},{ { \bm \omega}})=[{\mathcal U} v_2]({\bf x},{ { \bm
\omega}})$ implies that ${\bf z}^{(1)}\in
\partial G_0^{(2)}, \ {\bf z}^{(2)}\in \partial G_0^{(1)}$.
\end{theorem}

{Proof.}
	 Assume the opposite, that is, ${\bf z}^{(1)}\notin
\partial G_0^{(2)}$. Then ${\bf z}^{(1)}\in
G_0^{(2)}$ because $z^{(1)} \in \overline{G} =
\overline{G}^{(2)}_0 =
\partial{{G}^{(2)}_0} \bigcup {G}^{(2)}_0$.
Hence $(\Delta_{\bf x})^m[{\mathcal U}v_2]({\bf x},{ { \bm
\omega}})$ is continuous for ${\bf x}={\bf z}^{(1)}$, while the
function $(\Delta_{\bf x})^m[{ \mathcal U}v_1]({\bf x},{\bm
\omega})$ is discontinuous for ${\bf x}={\bf z}^{(1)}$, which
contradicts the condition of the theorem. That's why ${\bf
z}^{(1)}\in \partial G_0^{(2)}$. By symmetry of reasoning and
point ${\bf z}^{(2)}\in \partial G_0^{(1)}$. The theorem has been
proved.

     As can be seen, this statement has a local character. Let us also present the uniqueness theorem for the global
type, using the notation already introduced.

\begin{theorem} Let $V_1({\bf x},{\bf x})\neq 0$, $V_2({\bf x},{\bf x})\neq 0, \ {\bf x}\in
G$, and the functions $v_1({\bf y})$ and $v_2({\bf y})$ have
nonzero jumps in contact points of the surfaces $\partial
G_0^{(1)}$ and $\partial G_0^{(2)}$, respectively. Then from
coincidence $[{ \mathcal U} v_1]({\bf x},{ { \bm \omega}})$ and
$[{ \mathcal U}v_2]({\bf x},{ { \bm \omega}})$, $({\bf x},{ { \bm
\omega}})\in G\times {\Omega}$ the coincidence of $\partial
G_0^{(1)}$ and $\partial G_0^{(2)}$ follows.
\end{theorem}

{Proof.}
     As the surface  $\partial G$ is a common part of the surfaces $\partial
G_0^{(1)}$ and $\partial G_0^{(2)}$ then we need to compare only
the sets $\partial G_0^{(1)} \backslash\partial G$ and $\partial
G_0^{(2)} \backslash\partial G$.

It follows from theorem 2 that any contact point ${\bf z}
\in\partial G_0^{(1)} \backslash\partial G$ also belongs to
$\partial G_0^{(2)} \backslash\partial G$. Additionally any
contact point  ${\bf z} \in\partial G_0^{(2)} \backslash\partial
G$ also belongs to $\partial G_0^{(1)} \backslash\partial G$. This
means that the sets of the contact points in $\partial G_0^{(1)}
\backslash\partial G$ and in $\partial G_0^{(2)}
\backslash\partial G$ coincide. Hence, due to the density of
contact points  in the set $\partial G_0^{(1)} \backslash\partial
G$ and in the $\partial G_0^{(2)} \backslash\partial G$, we
conclude that $\partial G_0^{(1)} \backslash\partial G = \partial
G_0^{(2)} \backslash\partial G$. Therefore, $\partial
G_0^{(1)}=\partial G_0^{(2)}$. Theorem is proved.

\section{Conclusion}
      We  pay attention to the fact that the formula (\ref{7}) allows
us to construct an algorithm for solving the unknown boundary
problem studied in this paper. To do this, it suffices to perform
integration and differentiation of the given function and propose
any way to indicate the breakpoints of the received and already
known function on the left side of the formula (\ref{8}).


\end{document}